\documentclass[epj,nopacs]{svjour}
\usepackage{epsfig}
\usepackage{amssymb}
\usepackage{times}
\newcommand{\st}{{\scriptscriptstyle T}}
\def\F{{D^q_1(z,z^2 k_\st^2)}}
\def\Fs{{H_1^{\perp q}(z,z^2 k_\st^2)}}
\newcommand{\beq}  {\begin{equation}}
\newcommand{\eeq}  {\end{equation}}
\newcommand{\barr}{\begin{eqnarray}}
\newcommand{\earr}{\end{eqnarray}}
\begin{document}
\titlerunning{Transversity Distribution and Polarized Fragmentation
Function}
\title{Transversity Distribution and 
Polarized Fragmentation \\
Function from Semi-inclusive Pion Electroproduction}
\author{V.A.~Korotkov\inst{1, 2} \and W.--D.~Nowak\inst{2} 
\and K.A.~Oganessyan\inst{2, 3, 4}
}                     
\institute{IHEP, 142284 Protvino, Russia \and DESY Zeuthen, 
15738 Zeuthen, Germany 
\and INFN, Laboratori Nazionali di Frascati, 00044 Frascati, Italy 
\and Yerevan Physics Institute, 375036 Yerevan, Armenia}
\date{Received: date / Revised version: date}
%
\abstract{
A method is discussed to determine the hitherto unknown u-quark 
transversity distribution $\delta u(x)$ from a planned HERMES 
measurement of a single target-spin asymmetry in semi-inclusive pion 
electroproduction off a transversely polarized target.
Assuming u-quark dominance, the measurement yields the shapes of the 
transversity dis- \vspace*{-1.5ex}\\
tribution $\delta u(x)$ and of the ratio 
$H_1^{\perp (1)u}(z)/D_1^u(z)$, of polarized and unpolarized 
u-quark fragmentation functions. The unknown relative
normalization can be obtained by identifying the transversity
distribution with the well-known helicity distribution at large $x$
and small $Q^2$.
The systematic uncertainty of the method is dominated 
by the assumption of $u$-quark dominance.
\PACS{
      {PACS-key}{discribing text of that key}   \and
      {PACS-key}{discribing text of that key}
     } 
} 
\maketitle
%
\section{Introduction}
\label{sect1}
%
Deep inelastic charged lepton scattering off a {\it transversely} 
polarized nucleon target is an important tool to further 
study the internal spin structure of the nucleon. While a lot of 
experimental data on the longitudinal spin structure of the nucleon has 
been collected over the last 10 years, the study of its transverse 
spin structure is just about to begin. Only a very limited number 
of preliminary experimental results is available up to now: 
\begin{itemize}
\item [(1)] measurements of the nucleon structure function $g_2(x)$ 
  at CERN~\cite{smcg2} and SLAC~\cite{e154g2}-\cite{bosted99}, 
\item [(2)] a first measurement of a single target-spin asymmetry for 
  pions produced in lepton scattering off longitudinally polarized protons 
  at HERMES \cite{avakian}, 
\item [(3)] a first study of hadron azimuthal distributions in DIS of leptons 
  off a transversely polarized target at SMC \cite{bravar99}. 
\end{itemize}  

A quark of a given flavour is characterized by three twist-2 parton 
distributions. The quark number density distribution 
$q(x, Q^2)$ has been studied now for decades and is well known for all 
flavors.
The helicity distribution $\Delta q(x, Q^2)$ was 
only recently measured more accurately for $u-$ and $d-$quarks 
\cite{HERMESdeltaq} and is still essentially unknown for $s-$quarks. 
The third parton distribution, known generally as `transversity 
distribution' and denoted $\delta q(x, Q^2)$
characterizes the distribution of the quark's transverse spin in a 
transversely polarized nucleon. 

For non-relativistic quarks, where boosts and rotations commute,
$\delta q(x) = \Delta q(x)$. Since quarks in the nucleon 
are known to be relativistic, the difference between both 
distributions will provide further information on their relativistic 
nature. The trans\-versity distribution does not mix with gluons under 
QCD evolution, i.e. even if 
transversity and helicity distributions coincide at some scale, they will 
be different at $Q^2$ values higher than that.

The chiral-odd nature of transversity distributions makes their 
experimental determination difficult; up to now no experimental 
information on $\delta q(x, Q^2)$ is available. It can not be accessed 
in {\it inclusive} deep inelastic scattering (DIS) due to chirality 
conservation; it decouples from all hard processes that involve only one 
quark distribution (or fragmentation) function (see e.g. 
\cite{jaffe97b}). This is in contrast to the case of the 
chiral-even number density and helicity distribution
functions, which are directly accessible in inclusive lepton DIS.  

In principle, transversity distributions can be extracted from cross 
section asymmetries in polarized processes involving a transversely 
polarized nucleon. The corresponding asymmetry can be expressed 
through a flavor sum involving
products of two chiral-odd transversity distributions in the case of
hadron-hadron scattering, while in the case of {\it semi-inclusive} DIS 
(SIDIS) a chiral-odd quark distribution function always appears in 
combination with a chiral-odd quark fragmentation function. These 
fragmentation functions can in principle be measured in $e^+ e^-$ 
annihilation.  

The transversity distribution was first discussed by Ralston and Soper 
\cite{ralston} in doubly transverse polarized Drell-Yan scattering. Its 
measurement is one of the main goals of the spin program at RHIC 
\cite{rhic}. An evaluation of the corresponding
asymmetry $A_{TT}$ was carried out \cite{omartin} by
assuming the saturation of Soffer's inequality \cite{soffer95} 
for the transversity distribution. The maximum possible asymmetry at RHIC 
energies was estimated to be $A_{TT} = 1 \div 2 \%$. At smaller energies,
e.g. for a possible fixed-target hadron-hadron spin experiment 
HERA-$\vec N$ \cite{heran} ($\sqrt{s} \simeq 40$ GeV), the asymmetry is 
expected to be higher. 

In semi-inclusive deep inelastic lepton scattering off transversely 
polarized nucleons there exist several methods to access transversity 
distributions; all of them can in principle be realized at HERMES. One of
them, namely twist-3 pion production \cite{jaffe-ji93}, uses 
longitudinally polarized leptons and a double spin asymmetry is measured. 
The other methods do not require a polarized beam; they rely on 
{\it polarimetry} of the scattered transversely polarized quark:
\begin{itemize}
 \item [(1)] measurement of the transverse polarization of $\Lambda$'s in the 
       current fragmentation region \cite{baldracchini82}-\cite{jaffe96},
 \item [(2)] observation of a correlation between the transverse spin vector
       of the target nucleon and the normal to the two meson plane 
       \cite{jaffe97b,jaffe97a},
 \item [(3)] observation of the Collins effect in quark fragmentation through 
       the measurement of pion single target-spin asymmetries
       \cite{collins93}-\cite{mauro}.    
\end{itemize}          
The HERMES experiment \cite{Spectr} has excellent capabilities to investigate 
semi-inclusive particle production. 
Taking the measurement of the Collins effect as an example, 
it will be shown in the following that HERMES will be capable 
to extract both transversity and chiral-odd fragmentation function 
at the same time and with good statistical precision.

%
\section{Single Target-Spin Asymmetry in Pion Electroproduction}
\label{sect2}
%
A complete analysis of polarized 
SIDIS with non-zero transverse momentum effects in both the quark 
distribution and fragmentation functions was performed in the framework 
of the quark-parton model in \cite{kotzinian95} and in the field 
theoretical framework of QCD in \cite{mulders96}. An important 
ingredient of this analysis is the factorization property that was proven 
for $k_T$ integrated functions and that can reasonably be assumed for
$k_T$ depending functions \cite{mulders96}. In the situation that the 
final state polarization is not considered, two quark fragmentation 
functions are involved: $D^q_1(z,z^2 k_\st^2)$ and 
$H_1^{\perp q}(z,z^2 k_\st^2)$. Here $k_T$ is the intrinsic quark 
transverse momentum and $z$ is the fraction of quark momentum transfered
to the hadron in the fragmentation process.
The `polarized' fragmentation function $H_1^{\perp q}$ allows for 
a correlation between the transverse polarization of the fragmenting 
quark and the transverse momentum of the produced hadron. It may be 
non-zero because time reversal invariance is not applicable in a decay
process, as was first discussed by Collins~\cite{collins93}.

Since quark transverse momenta cannot be measured directly, 
integrals over $k_\st$ (with suitable weights) are defined to arrive at
experimentally accessible fragmentation functions:
\barr
 z^2\int d^2k_\st\,\F  & \equiv & D^q_1(z)
\earr
is the familiar unpolarized fragmentation function, normalized by the 
momentum sum rule $\sum_h \int dz\,zD_1^{q\rightarrow h}(z) = 1$.
Correspondingly, the polarized fragmentation function is obtained as
\barr
z^2\int d^2 k_\st \,\left(\frac{k_\st^2}{2M_h^2}\right)
\,H_1^{\perp q}(z,z^2k_\st^2) & \equiv & H_1^{\perp (1)q}(z),
\earr
where the superscript $(1)$ indicates that an originally $k_T$ depen\-dent 
function was integrated over $k_T$ with the weight 
$\frac{k_\st^2}{2M_h^2}$. Here, $M_h$ is the mass of the produced hadron $h$.  

To facilitate access to transversity and polarized fragmentation 
functions from SIDIS, single-spin asymmetries may be formed through
integration of the polarized cross section over $P_{h\perp}$,
the transverse momentum of the final hadron, with
appropriate weights. In the particular case of an unpolarized beam 
and a transversely polarized target the following {\it weighted asymmetry} 
provides access to the quark transversity distribution via the Collins 
effect \cite{kotmul97}:
\barr
\lefteqn{A_T(x,y,z) \equiv}  \nonumber \\ 
& &\frac{\int d \phi^\ell \int d^2P_{h\perp}\, 
\frac{\vert P_{h\perp}\vert}{zM_h}
\sin(\phi_s^\ell + \phi_h^\ell)
\,\left(d\sigma^{\uparrow}-d\sigma^{\downarrow}\right)}
{\int d \phi^\ell \int d^2 P_{h\perp} (d\sigma^{\uparrow}+d\sigma^{\downarrow})}.
\earr
Here $\uparrow (\downarrow)$ denotes target up (down) transverse polarization.
The azimuthal angles are defined in the transverse
space giving the orientation of the lepton plane ($\phi^\ell$) and the
orientation of the hadron plane ($\phi^\ell_h$ = $\phi_h - \phi^\ell$) 
or spin vector ($\phi^\ell_s$ = $\phi_s - \phi^\ell$)
with respect to the lepton plane. The angles are measured around 
the z-axis which is defined by the momenta $q$ and $P$ of the virtual 
photon and the target nucleon, respectively. 
The raw asymmetry (3) can be estimated \cite{kotmul97} using
\beq
\label{asymmetry}
A_T(x,y,z) =
P_T \cdot D_{nn} \cdot
\frac{\sum_q e^2_q \  \delta q(x) \ H_1^{\perp(1)q}(z)}
     {\sum_q e^2_q \  q(x) \ D^q_1(z)} ,
\eeq
where $P_T$ is the target polarization and
$D_{nn} = (1-y)/(1-y+y^2/2)$ is the transverse spin transfer coefficient.
The magnitude of the asymmetry depends on the unknown
functions $\delta q(x)$ and $H_1^{\perp(1)q}(z)$.  

\section{Transversity Distribution and Polarized Fragmentation
Function}
\label{sect3}
\begin{figure}[h]
\centering
\epsfig{file=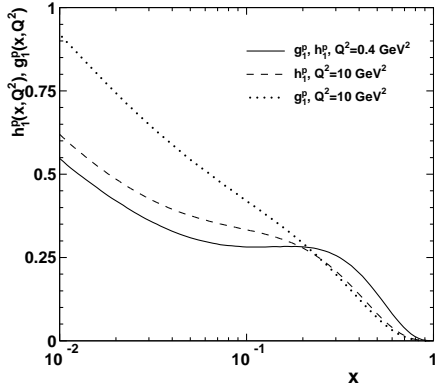,width=6.0cm}
\caption{\it The transversity distribution $h_1^p(x,Q_0^2)$ 
(continuous line) which coincides with the helicity distribution 
$ g_1^p(x,Q_0^2)$  at the scale $Q_0^2 =0.4\ GeV^2$ (as given
by the GRSV LO parameterization \cite{grsv96} in the 'standard' scenario).
Their evolved LO distributions $h_1^p(x,Q^2)$ (dotted) and 
$g_1^p(x,Q^2)$ (dot-dashed) are shown at $Q^2= 10\ GeV^2$. 
}
\label{h1evol}
\end{figure}

No experimental data are available on any of the transversity
distributions $\delta q(x)$, while their behaviour under 
QCD-evolu\-tion is theoretically well established~\cite{ArtruMek}.
An example for the leading order evolution of  the proton
structure functions 
$g_1(x, Q^2)$ $= \frac{1}{2} {\sum_i e_i^2 \Delta q_i(x,Q^2)}$ 
and $h_1(x, Q^2) = \frac{1}{2} {\sum_i e_i^2 \delta q_i(x,Q^2)}$ is shown in 
Fig.~\ref{h1evol}. It was assumed that $h_1^p(x)$ coincides with $g_1^p(x)$ at 
the scale $Q_0^2 \, = \, 0.4$~GeV$^2$ and both functions were evolved to the
scale $Q^2 \, = \, 10$~GeV$^2$. The evolution
was performed using
the programs from \cite{kumanog1,kumanoh1} for $g_1$
and $h_1$, respectively. The important conclusion, which was already 
discussed earlier (see e.g. \cite{scopetta}), follows that with 
increasing $Q^2$ the two functions are becoming more and more 
different for decreasing $x$ while at large $x$ the difference remains 
quite small.

Results from two independent measurements indicate that the polarized 
fragmentation function $H_1^{\perp(1)q}(z)$ may be non-zero: 
i) azimuthal correlations measured between particles produced from 
opposite jets in $Z$ decay at DELPHI \cite{delphi2} and ii) the single 
target-spin asymmetry measured for pions produced in SIDIS of leptons
off a longitudinally polarized target at HERMES \cite{avakian}. 
The approach of \cite{kotmul97} is adopted to estimate the possible 
value of $H_1^{\perp(1)q}(z)$. 
Collins~\cite{collins93} suggested the following parameterization for the 
analyzing power in transversely polarized quark fragmentation:
\beq
A_C(z,k_T) \equiv \frac{\vert k_T\vert\Fs}{M_h\F}
=\frac{M_C\,\vert k_T\vert}{M_C^2+\vert k_T^2\vert },
\eeq
with $M_C \simeq 0.3\div 1.0$~GeV being a typical hadronic mass. Choosing
a Gaussian parameterization for the quark transverse momentum dependence
in the unpolarized fragmentation function
\beq
\F=D^q_1(z)\,\frac{R^2}{\pi\,z^2}\,\exp(-R^2 k_T^2),
\eeq
leads to
\barr
\lefteqn{H_1^{\perp (1)q}(z) =} \nonumber \\
& &D^q_1(z)\frac{M_C}{2M_h}\left(1-M_C^2R^2
\int_{0}^{\infty}dx\,\frac{\exp(-x)}{x+M_C^2R^2}\right).
\earr
Here $R^2=z^2 / b^2$, and $b^2$ is the mean-square momentum the hadron 
acquires in the quark fragmentation process. In the following the parameter
settings $M_C = 0.7$ GeV and $b^2 = 0.25$ GeV$^2$ are used because they
are consistent \cite{kotzinian99} with the single target-spin asymmetry 
measured at HERMES \cite{avakian}. They are also compatible with the 
analysis of \cite{delphi2}, as can be seen by evaluating the ratio 
\begin{equation}
\label{ffratio}
R(z_{min}) = {{\int_{z_{min}}^1 \, dz H_1^{\perp} (z) } \over
              {\int_{z_{min}}^1 \, dz D_1 (z) } } ,
\end{equation}
where $H_1^{\perp} (z)$, in contrast to Eq.(2), is the  
unweighted polarized fragmentation function used in \cite{delphi2}
\begin{equation}
z^2\int d^2 k_\st
\,H_1^{\perp q}(z,z^2k_\st^2) \, \equiv \, H_1^{\perp q}(z).
\end{equation}
The BKK parameterization \cite{bkk} was used to estimate the integral over 
the unpolarized fragmentation function $D_1(z)$. The values obtained for the 
ratio, $R( 0.1 ) = 0.048$ and $R( 0.2 ) = 0.070$, are to be compared to 
the experimental result~\cite{delphi2}: $0.063 \pm 0.017$.  

%
\section{Projected Statistical Accuracy and Systematics}
\label{sect4}
%

A full analysis to extract transversity and polarized fragmentation
functions through (4) requires one to take into account
all quark flavours contributing to the measured asymmetry. According to 
calculations with the HERMES Monte Carlo program HMC, the fraction of positive 
pions originating from the fragmentation of a struck u-quark ranges,
depending on the value of $x$,  between $70$ and $90\%$ 
for a proton target and is
only slightly smaller for a deuteron target. Therefore, in a first
analysis, the assumption of $u$-quark dominance in the $\pi^+$ production
cross-section appears to be reasonable.
This is supported by the sum rule for T-odd fragmentation
functions recently derived in \cite{schaefer}.
These authors concluded that contributions from non-leading parton 
fragmentation,
like $d \rightarrow \pi^+$, is severely suppressed for all T-odd
fragmentation functions.

Consequently, the assumption of $u$-quark dominance 
was used to calculate projections for 
the statistical accuracy in measuring the asymmetry 
$A_T^{\pi^+}(x)$. The expected statistics for scattering at HERMES
unpolarized leptons off a transversely polarized target (proton or deuteron
options are under consideration) will
consist of about seven millions reconstructed DIS events.
The standard definition of a DIS event at HERMES is given by 
the following set of kinematic 
cuts\footnote{$Q^2$ and $\nu$ are the photon's virtuality and 
laboratory energy, $x = Q^2 / 2 M \nu$ is the Bjorken scaling variable, 
$y = \nu/E$ is the fractional photon energy and $W$ is the c.m. energy
of the photon-nucleon system; $E = 27.5$~GeV at HERMES.}:
\begin{center}
$Q^2  \, > 1 $~GeV$^2$, $W  \, > 2 $~GeV, 
$0.02 \, < \,  x  \, < \, 0.7$, $y  \, < \, 0.85$.  
\end{center}
An additional cut $W^2 > 10$~GeV$^2$ was introduced in the analysis
to improve the separation of the struck quark fragmentation region. 
An average target polarization of $P_T = 75 \%$ is used for the analysis. \\
Considering only $u$-quarks the expression for the asymmetry 
(\ref{asymmetry}) reduces to the simple  form
\beq
\label{uasymmetry}
A_T(x,y,z) =
P_T \cdot D_{nn} \cdot
\frac{\delta u(x)}{u(x)} \cdot \frac{H_1^{\perp(1)u}(z)}{D^u_1(z)} 
\eeq
for a proton target, and
\barr
\label{dasymmetry}
\lefteqn{A_T(x,y,z) =} \nonumber \\
& & ( 1 - \frac{3}{2} \omega_D) \cdot P_T \cdot D_{nn} \cdot
\frac{\delta u(x) + \delta d(x)}{u(x) + d(x)} \cdot 
\frac{H_1^{\perp(1)u}(z)}{D^u_1(z)} \; \; \;
\earr
for a deuteron target. Here $\omega_D = 0.05 \pm 0.01$ is the probability
of the deuteron to be in the $D$-state. \\
To simulate a measurement of $A_T$ the approximation 
$\delta q(x)  =  \Delta q(x)$ could be used in view of the relatively 
low $Q^2$-values at HERMES, in accordance with the above discussion.
The Gehrmann-Stirling parameterization in leading order \cite{gehrstir} 
was taken for $\Delta q(x)$ and the GRV94LO parameterization \cite{grv94lo} 
for $q(x)$. The $Q^2$ evolution of the quark distributions was neglected 
and  $Q^2 = 2.5$ GeV$^2$ was taken as an average value for the HERMES 
kinematical region. The HERMES Monte Carlo program HMC was used to account 
for the spectrometer acceptance. The following cuts were applied to the 
kinematic variables of the pion\footnote{$x_F = 2 p_L / W$, where $p_L$ is the
longitudinal momentum of the hadron with respect to the virtual photon in the
photon-nucleon c.m.s., and $z = E_h / \nu$, where $E_h$ is the energy of
the produced hadron.}: 
\begin{center}
$x_F > 0$, $z > 0.1$, $P_{h \perp} > 0.05$ GeV .
\end{center}

\begin{figure}[htb]
\centering
\epsfig{file=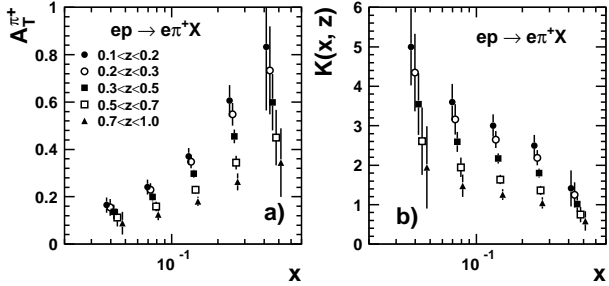,width=9cm}
\caption{\it Proton target. {\bf a)} The weighted asymmetry $A_T^{\pi^+}(x)$ 
in different intervals of $z$; {\bf b)} the function $K(x, z)$.  }
\label{collins1}
\end{figure}

The simulated data were divided into 5x5 bins in ($x,z$). The expectations 
for the asymmetry $A_T^{\pi^+} (x)$ as would be measured by HERMES
using a proton target, are presented in Fig.~\ref{collins1}a
in different intervals of the pion variable $z$.
The projected accuracies for the asymmetry were estimated according to
\beq
 \delta A_T = \left \langle \Bigl ({{P_{h\perp}}\over{{z m_\pi}}} 
                           \sin(\phi_s^\ell + \phi_h^\ell) \Bigr ) ^2 
\right \rangle ^{\frac{1}{2}} \cdot {{1}\over{\sqrt{N_\pi}}} ,
\eeq
where $N_{\pi}$ is the total number of measured positive pions after 
kinematic cuts, and $< \cdots >$ means averaging over all accepted events. 
In this way the product of the transversity 
distribution and the ratio of the fragmentation functions,
\beq
 K(x, z) = \delta u(x) \cdot \frac{H_1^{\perp(1)u}(z)}{D^u_1(z)},
\eeq
as well as the projected statistical accuracy for a measurement of 
this function were calculated and are shown in Fig.~\ref{collins1}b,
again for the case of a proton target.

\begin{figure}[htb]
\centering
\epsfig{file=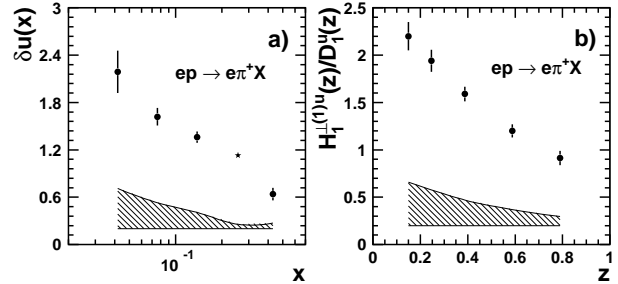,width=9cm}
\caption{\it
 {\bf a)} The transversity distribution $\delta u(x)$, and
{\bf b)} the ratio of the fragmentation functions $H_1^{\perp (1) u} (z)$ and
$D_1^u (z)$ as would be measured by HERMES with a {\rm proton} target. 
The asterisk in {\bf a)} shows the normalization point.
}
\label{collins2}
\end{figure}

The factorized form of expression (\ref{uasymmetry}) with respect to the 
variables $x$ and $z$ allows the simultaneous reconstruction
of the shape for the two unknown functions $\delta u(x)$ and
 $H_1^{\perp(1)u}(z) / D^u_1(z)$, while the relative normalization
cannot be fixed without a further assumption. As was discussed above,
the transversity distribution $\delta q(x)$ conceivably coincides with 
the helicity distribution $\Delta q(x)$ at small values of $Q^2$
where the relativistic effects are expected to be small. According to 
Fig.~\ref{h1evol} the differences are smallest in the region of
intermediate and large values of $x$. Hence the assumption 
\begin{equation}
\label{h1eqg1}
\delta q (x_0) = \Delta q(x_0)
\end{equation}
at $x_0=0.25$ was made to resolve the normalization ambiguity. The 
experimental data then consist of 25 measured values of the function 
$K(x_i, z_j)$, as opposed to 9 unknown function values: 
4 values for $\delta u(x_i)$ and 5 values for
 $H_1^{\perp(1)u}(z_j) / D^u_1(z_j)$, where the indices $i$ and $j$ 
enumerate the experimental intervals in $x$ and in $z$, respectively.
The standard procedure of $\chi^2$ minimization was applied to 
reconstruct the values for both $\delta u(x)$ and  
$H_1^{\perp(1)u}(z) / D^u_1(z)$ and to evaluate their projected 
statistical accuracies expected for a real measurement at HERMES. The
results are shown in Figs.~\ref{collins2}a,b, 
respectively.

\begin{figure}[htb]
\centering
\vspace*{-1.0cm}
\epsfig{file=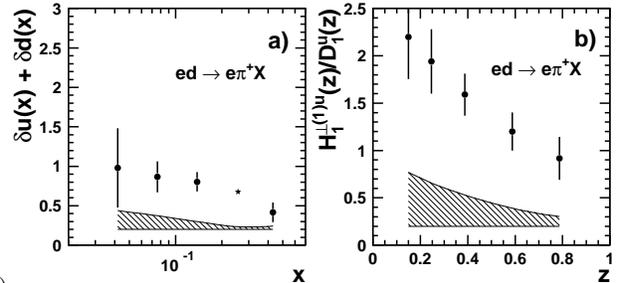,width=9cm}
\caption{\it
 {\bf a)} The transversity distribution $\delta u(x) + \delta d(x)$, and
{\bf b)} the ratio of the fragmentation functions $H_1^{\perp (1) u} (z)$ and
$D_1^u (z)$ as would be measured by HERMES with a {\rm deuteron} target. 
The asterisk in {\bf a)} shows the normalization point.
}
\label{deuteron}
\end{figure}

In an analogous way the consideration of the deuteron 
asymmetry (\ref{dasymmetry}) 
allows the evaluation of the projected statistical
accuracies for a measurement of the functions $\delta u(x) + \delta d(x)$ and
$H_1^{\perp (1) u} (z) / D_1^u (z)$ (see Figs.~\ref{deuteron}a,b,
respectively). The projected statistical accuracy 
is considerably worse than that for the proton target;
this is caused mainly by the expected smaller value of the asymmetry 
(\ref{dasymmetry}), which in turn is due to the lower value of 
$(\delta u(x) + \delta d(x)) / (u(x) + d(x))$ compared to
$\delta u(x) / u(x)$.


\begin{figure}[htb]
\centering
\epsfig{file=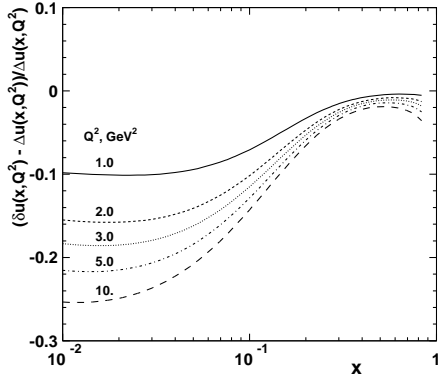,width=6cm}
\caption{\it
Relative difference between transversity distribution $\delta u(x, Q^2)$ and
helicity distribution $\Delta u(x, Q^2)$ as a function of $x$ and $Q^2$
in the kinematical region accessible to the HERMES experiment
($< Q^2 > \simeq 2.5$~GeV$^2$).
}
\label{ghdiff}
\end{figure}

Two sources of systematic uncertainties arising from approximations used
in the analysis were investigated. To evaluate the contribution of the
normalization assumption (\ref{h1eqg1}), the relative difference between 
transversity distribution $\delta u(x, Q^2)$ and helicity distribution 
$\Delta u(x, Q^2)$ was studied as a function of $x$ and $Q^2$ in the 
HERMES kinematics. Starting from $\delta u(x) = \Delta u(x)$ at the scale 
$Q_0^2 \, = \, 0.4$~GeV$^2$ both functions 
were evolved to higher values of $Q^2$. The results are shown in 
Fig.~\ref{ghdiff} and allow the conclusion that the relative difference 
is small for $x$ above $0.2 \div 0.3$; the corresponding systematic 
uncertainty is on the level of $2 \div 5$\%. 
The same conclusion is valid for the  evolution of 
$\delta u(x) + \delta d(x)$. 
A larger contribution to 
the systematic uncertainty originates from the above mentioned 
`contamination' of other quark flavors than $u$ to $\pi^+$ production,
when assuming u-quark dominance in the analysis. The $x$ and 
$z$ dependence of this contamination was evaluated with HMC. Both 
contributions were added linearly; the resulting total projected
systematic uncertainties on the extraction of the transversity distribution 
$\delta u(x)$  and the fragmentation function ratio 
$H_1^{\perp(1)u}(z) / D_1^u (z)$,
as would be measured using a proton target,
are shown as hatched bands in Figs.~\ref{collins2}a,b, as a function 
of $x$ and $z$, respectively.  
The same procedure for a deuteron target yields projected systematic
uncertainties for $\delta u(x) + \delta d(x)$ and 
$H_1^{\perp(1)u}(z) / D_1^u (z)$, 
as shown as hatched bands in Figs.~\ref{deuteron}a,b, respectively. 

%
\section{Conclusions}
\label{sect5}
%

In conclusion, the HERMES experiment
using a transversely polarized proton target
will be capable to 
measure simultaneously and with good statistical precision the shapes of the
u-quark transversity distribution $\delta u(x)$ and of the 
ratio of the fragmentation functions $H_1^{\perp (1) u}(z) / D_1^u (z)$.
The normalization can be fixed under the assumption 
that in the HERMES $Q^2$ range the transversity distribution is well 
described by the helicity distribution at large $x$. 
Using a deuteron target, information on 
$\delta u(x) + \delta d(x)$ will be available,
but with considerably less statistical accuracy 
compared to a measurement of $\delta u(x)$ from a proton target.
The systematic 
uncertainty of the method proposed in this paper is dominated by the 
assumption of $u$-quark dominance.   \\

%
\section*{Acknowledgements}
%
We thank Klaus Rith for discussions which led to the development of the 
method described in this paper and
the HERMES Collaboration for the kind permission to use their Monte Carlo
Program. We are indebted to Ralf Kaiser for the careful reading of the
manuscript.


\begin{thebibliography}{}
\bibitem{smcg2} 
SMC Collaboration, D.~Adams et al., Phys. Rev. \textbf{D56}, 5330 (1997).
\bibitem{e154g2}
E154 Collaboration, K.~Abe et al.,  Phys. Lett. \textbf{B404}, 377 (1997).
\bibitem{e143g2}
E143 Collaboration, K.~Abe et al., Phys. Rev. \textbf{D58}, 112003 (1998).
\bibitem{bosted99}
P.~Bosted, {\it Very Preliminary Results for the Spin Structure
Function $g_2$ from SLAC E155x}, Proc. of 15th Int. Conf. on Part. 
and Nuclei (PANIC 99), Uppsala, Sweden, June 1999.
\bibitem{avakian}
HERMES Collaboration, A.~Airapetian et al., Phys. Rev. Lett. \textbf{84}, 
4047 (2000).
\bibitem{bravar99}
A. Bravar, Nucl. Phys. Proc. Suppl. \textbf{79}, 520 (1999).
\bibitem{HERMESdeltaq}
HERMES Collaboration, K.~Ackerstaff et al., Phys. Lett. \textbf{B464},
123 (1999).
\bibitem{jaffe97b}
R.L.~Jaffe, Proc. of the 2nd Topical Workshop
{\it Deep Inelastic Scattering off Polarized Targets: Theory Meets Experiment},
DESY 97-200, Zeuthen, 1997, p.167, ed. by J.~Bl\"umlein, A.~de~Roeck,
T.~Gehrmann, and W.-D.~Nowak; hep-ph/9710465.
\bibitem{ralston}
J. Ralston, D.E. Soper, Nucl. Phys. \textbf{B152}, 109 (1979).
\bibitem{rhic}
D.~Hill et al., RHIC Spin Collaborationboration: Letter of Intent, BNL, 
May 1991.
\bibitem{omartin}
O. Martin et al., Phys. Rev. \textbf{D60}, 117502 (1999).
\bibitem{soffer95}
J.~Soffer,  Phys. Rev. Lett. \textbf{74}, 1292 (1995).
\bibitem{heran}
V.A. Korotkov and W.-D. Nowak, Proc. of the Workshop {\it Polarized Protons 
at High Energies - Accelerator Challenges and Physics Opportunities},
DESY-PROC-1999-03, 1999, p.19, ed. by A.de~Roeck, D.~Barber, and G.~R\"adel; 
DESY 99-122; hep-ph/9908490.
\bibitem{jaffe-ji93}
R.L.~Jaffe, X.J.~Ji, Phys. Rev. Lett. \textbf{71}, 2547 (1993).
\bibitem{baldracchini82}
F.~Baldracchini et al., Fortschritte der Phys. \textbf{29}, 505 (1981).
\bibitem{ArtruMek}
X.~Artru, M.~Mekhfi, Zeit. Phys. \textbf{C45}, 669 (1990).
\bibitem{jaffe96}
R.L.~Jaffe, Phys. Rev. \textbf{D54}, 6581 (1996).
\bibitem{jaffe97a}
R.L.~Jaffe, X.~Jin, J.~Tang, Phys. Rev. Lett. \textbf{80}, 1166 (1998).
\bibitem{collins93}
J.C.~Collins, Nucl. Phys. \textbf{B396}, 161 (1993).
\bibitem{collins94}
J.C.~Collins, S.F.~Heppelmann, G.A.~Ladinsky, 
Nucl. Phys. \textbf{B420}, 565 (1994).
\bibitem{kotzinian95}
A.~Kotzinian, Nucl. Phys. \textbf{B441}, 234 (1995).
\bibitem{mulders96}
P.J.~Mulders, R.D.~Tangerman, Nucl. Phys. \textbf{B461}, 197 (1996); \\
Erratum - ibid. \textbf{B484}, 538 (1997).
\bibitem{mauro}
M.~Anselmino, E.~Leader, F.~Murgia, Phys. Rev. \textbf{D56}, 6021 (1997).
\bibitem{Spectr}
K.~Ackerstaff et al., Nucl. Instr. and Methods \textbf{A417}, 230 (1998).
\bibitem{kotmul97}
A.M.~Kotzinian, P.J.~Mulders, Phys.Lett. \textbf{B406}, 373 (1997).
\bibitem{grsv96}
M. Gl\"uck et al., Phys. Rev. \textbf{D53}, 4775 (1996).
\bibitem{kumanog1}
M. Hirai, S. Kumano, M. Miyama, Comp. Phys. Comm. \textbf{108}, 38 (1998).
\bibitem{kumanoh1}
M. Hirai, S. Kumano, M. Miyama, Comp. Phys. Comm. \textbf{111}, 150 (1998).
\bibitem{scopetta}
S.~Scopetta, V.~Vento, Phys. Lett. \textbf{B424}, 25 (1998).
\bibitem{delphi2}
A.V. Efremov, O.G. Smirnova, L.G. Tkachev,
Nucl. Phys. Proc. Suppl. \textbf{74}, 49 (1999). 
\bibitem{kotzinian99}
A.M. Kotzinian et al., hep-ph/9908466.
\bibitem{bkk}
J.~Binnewies, B.A.~Kniehl, G.~Kramer, Zeit. Phys. \textbf{C65}, 471 (1995).
\bibitem{schaefer}
A.~Sch\"afer, O.~Teryaev, Phys. Rev. \textbf{D61}, 077903 (2000).
\bibitem{gehrstir}
T.~Gehrmann, W.J.~Stirling, Phys. Rev. \textbf{D53}, 6100 (1996).
\bibitem{grv94lo}
M.~Gl\"uck, E.~Reya, A.~Vogt, Zeit. Phys. \textbf{C67}, 433 (1995).
\end{thebibliography}
\end{document}